\documentclass[sensors,article,accept,moreauthors,pdftex,AddToHook]{Definitions/mdpi} 
 
\usepackage{gensymb}
\usepackage{siunitx}
\usepackage{xcolor}
\usepackage{graphicx}
\usepackage[normalem]{ulem}
\usepackage{caption}
\usepackage{subcaption}
\usepackage[none]{hyphenat}

\usepackage[acronym,toc,shortcuts]{glossaries}

\usepackage[normalem]{ulem}
\newif\ifrev
\revtrue 
\ifrev

\newcommand{\rmtexto}[1]{\textcolor{red}{\sout{{#1}}}}

\else

\newcommand{\rmtexto}[1]{}
\fi

\newacronym{eeg}{EEG}{electroencephalogram}
\newacronym{who}{WHO}{World Health Organization}
\newacronym{at}{AT}{Assistive Technology}
\newacronym{bci}{BCI}{Brain-Computer Interface}
\newacronym{als}{ALS}{Amyotrophic Lateral Sclerosis}
\newacronym{ml}{ML}{Machine Learning}
\newacronym{uav}{UAV}{Unmanned Aerial Vehicles}
\newacronym{cns}{CNS}{Central Nervous System}
\newacronym{pns}{PNS}{Peripheral Nervous System}
\newacronym{cnn}{CNN}{Convolutional Neural Network}
\newacronym{els-iin}{ELS-IIN}{Edmond and Lily Safra International Institute of Neurosciences}

\firstpage{1} 
\pubvolume{1}
\issuenum{1}
\articlenumber{0}
\pubyear{2023}
\copyrightyear{2023}
\externaleditor{Academic Editors: Georg Fischer and Sung-Phil Kim}
\datereceived{ } 
\daterevised{ } 
\dateaccepted{ } 
\datepublished{ } 
\hreflink{https://doi.org/} 

\Title{Electroencephalography Signal Analysis for Human Activities Classification: A Solution Based on Machine Learning and Motor Imagery}
\TitleCitation{Electroencephalography Signal Analysis for Human Activities Classification: A Solution Based on Machine Learning and Motor Imagery}

\Author{Tarciana C. de Brito Guerra 
 $^{1,}$*\orcidA{}, Taline Nóbrega $^{1}$\orcidB{}, Edgard Morya  $^{2}$\orcidC{}, Allan de M. Martins $^{1}$\orcidD{} and~\mbox{Vicente A. de 
 Sousa Jr.}$^{1}$\orcidD{}}

\AuthorNames{Tarciana C. de Brito Guerra, Taline Nóbrega, Edgard Morya, Allan de M. Martins and Vicente A. de Sousa Jr.}
\AuthorCitation{Brito Guerra, T.C.; Nóbrega, T.; Morya, E.; Martins, A. M.; de Sousa, V.A., Jr.}

\address{%
$^{1}$ \quad Graduate Program in Electrical and Computer Engineering (PPgEEC), Federal University of Rio Grande do Norte, Natal 59078-970, Brazil; taline.nobrega@gmail.com (T.N.); allan@dca.ufrn.br (A.M.M.); vicente.sousa@ufrn.br (V.A.S.J.)\\
$^{2}$ 
\quad Graduate Program in Neuroengineering, Edmond and Lily Safra International Institute of Neuroscience, Santos Dumont Institute, Macaíba 59280-000, Brazil; edgard.morya@isd.org.br (E.M.)
}

\corres{Correspondence: tarciana.guerra.051@ufrn.edu.br (T.C.B.G)} 

\abstract{Electroencephalography (EEG) is a fundamental tool for understanding the brain's electrical activity related to human motor activities. Brain-Computer Interface (BCI) uses such electrical activity to develop assistive technologies, especially those directed at people with physical disabilities. However, extracting signal features and patterns is still complex, sometimes delegated to machine learning (ML) algorithms. Therefore, this work aims to develop a ML based on the Random Forest algorithm to classify EEG signals from subjects performing real and imagery motor activities. The interpretation and correct classification of EEG signals allow the development of tools controlled by cognitive processes. We evaluated our ML Random Forest algorithm using a consumer and a research-grade EEG system. Random Forest efficiently distinguishes imagery and real activities and defines the related body part, even with consumer-grade EEG. However, interpersonal variability of the EEG signals negatively affects the classification process.}


\begin{document}

\section{Introduction}
\label{sec_intro}

The \ac{who} published the Global Disability Action Plan in 2015~\cite{WHO_plan}. The document presents statistical data and establishes objectives and action plans to promote quality of life for people with some disabilities. The main goals involve removing barriers and improving access to health services and programs. People with disabilities represent about 15\% of the world population, which will increase due to longer life expectancy and continuous technological advances~\cite{WHO_plan}. 

The development of assistive technologies promotes accessibility and rehabilitation~\cite{ONA_report}. The \ac{who} on Disability ~\cite{ONA_report} suggests research actions for developing technological devices to provide accessibility and rehabilitation. These devices could be a simple crutch, cane, or even a more complex solution, such as \ac{bci}.

The advances in neuroscience associated with digital signal processing and biomedical instrumentation allow a better understanding of the brain's functions \cite{thesis_cleison}. These studies enable the development of systems capable of translating biological signals into digital responses~\cite{shih}. Hans Berger recorded the first brain electrical activity from a human scalp~\cite{graimann}. Years later, Grey Walter described the first \ac{bci}. He implanted electrodes directly into the motor cortex and recorded brain activity. The results showed the possibility of predicting body movement by analyzing the brain's signal~\cite{graimann, rotta}. 


The results of these researches stimulated interest in \ac{bci}, as they led to the possibility of implementing systems capable of helping impaired people, such as people with quadriplegia and some degenerative disease, for example, \ac{als}~\cite{wolpaw}. BCI is also used in other proposals, such as the control of \ac{uav} \cite{lafleur}. Therefore, its relevance is not limited to applications for people with physical and motor \hl{limitations} \cite{tese_Yule}. 

\ac{bci} systems use \ac{eeg} to acquire brain electrical activity through non-invasive electrodes and safe and low-cost procedures \cite{thesis_cleison, wolpaw}. This method has an excellent temporal resolution, allowing applications in many \hl{fields} \cite{tese_Werley}.

Extracting information robustly and reliably from \ac{eeg} signals can be challenging \cite{bashashati}. Most projects generally consider the properties of \ac{eeg} signals in the time, frequency, and space domains \cite{rodriguez-Bermude}. The objective is to identify the features of the signals associated with the brain activity of interest so that it is possible to control a \ac{bci} based on the user's intentions. Thus, processing these signals considers two fundamental steps: feature extraction and classification \cite{thesis_cleison}.

Another problem involving \ac{bci} is rapid data manipulation; the user must experience real-time interactions that mimic the body's natural actions \cite{tese_Ivo}. The natural interaction with the external environment occurs from communicating the \ac{cns} with the muscles through the \ac{pns}. This communication is efferent - from \ac{cns} to \ac{pns} (control of voluntary movement); and afferent---from the sensory system to the \ac{cns} (sensation and perception). Any damage to these pathways can result in sensorimotor problems and permanent damage to motor \hl{control} \cite{tese_Yule}.

\ac{eeg} is widely used to control \ac{bci}. However, these signals are very susceptible to noise, and pattern identification is complex \cite{tese_Ivo}. In this context, it is possible to use data analysis to interpret these signals adequately. Studies show that machine learning algorithms efficiently explore and classify neural activities~\cite{craik2019}. These algorithms automatically classify the data, allowing more practical applications, and it is less dependent on trained professionals~\cite{craik2019}.

The acquisition protocols used to collect \ac{eeg} signals usually involve motor tasks, sensory stimuli, such as image and sound presentation, and activities that comprise mental simulation of a motor act, also known as motor \hl{imagery} \cite{tese_Yule}. The protocol depends on the application of interest. \ac{bci} implemented for people with some motor disability generally aims to implement acquisition protocols independent of motor control. The use of motor imagery makes it possible, for example, for a person with paraplegia to control a neuroprosthesis since the signal to be collected does not depend on the remaining muscle activity~\cite{tese_mariakaroline}.

Thus, applications with this type of protocol not only benefit the control of devices through \ac{eeg} signals but also allow the rehabilitation of people with motor disabilities~\cite{tese_mariakaroline}. This type of rehabilitation is possible due to neuronal plasticity, which is the ability of neurons to establish new connections from new stimuli, which may come from the organism itself or the external environment \cite{wolpaw}. The brain can learn new functions and adapt itself, even when some regions have functional impairment \cite{tese_mariakaroline}.

The rest of this paper is structured as follows. Section \ref{sec_works} shows the research background and the related works, while Section \ref{sec_sol} focuses on our proposed solution, including the details of signal processing and \ac{ml} structure. Section \ref{sec_setup} gives information on the experiment setup and the measurement protocol, and Section \ref{sec_results} shows the results. Finally, Section \ref{sec_conclusions} describes our conclusions and final remarks.

\section{Related Works and Research Background}
\label{sec_works}
The technological evolution of the last decades allowed the development of algorithms capable of solving highly complex tasks, such as those related to \ac{eeg} signal processing. The information from such signals benefits several applications, not restricted to people with motor limitations. The control of \ac{bci} from interpreting and analyzing these signals has aroused increasing interest in the scientific community. Machine learning studies to solve paradigms involving BCI and EEG have increased classification efficiency. The challenge is to develop algorithms capable of providing high reliability and slight response delay concerning pattern recognition of \ac{eeg} data.

Amin et al. \cite{amin2019} pointed out some challenges inherent to applications involving \ac{eeg} signal decoding of motor imagery tasks. They proposed the fusion of several models of \ac{cnn}. The method considered multi-layer \ac{cnn}s with different architectures and characteristics to improve classification accuracy. Also studying \ac{eeg} signals in applications with motor imagery, Sadiq et al. \cite{Sadiq2019} developed a signal decomposition method based on empirical Wavelet Transform. The study involved the analysis of 18 electrodes to investigate the non-stationary and non-linear behavior of \ac{eeg} signals, considering the power spectral density and the Hilbert Transform. The researchers considered the power spectral density and the Hilbert Transform. The classification accuracy was 95.2~$\%$, and could be better by incorporating higher-order statistical resources, such as kurtosis.

Craik et al. \cite{craik2019} presented a systematic literature review involving task classification using electroencephalographic signals with machine learning tools. The research revealed that approximately $40\%$ of tasks with motor imagery protocols used \ac{cnn} as learning architecture. However, the research did not show the use of Random Forest as a classification technique. Liu et al. \cite{liu2017} used Random Forest with motor actions of the lower limbs, aiming to describe the impact of different feedback modalities of the \ac{bci} system based on EEG signals that decode extension and flexion of the legs. The results considered real-time simulations and presented average accuracy of 62.33$\%$ and 63.89$\%$ for two sessions.

Lazurenkoa et al. \cite{lazurenkoa2019}, and Meziani et al. \cite{meziani2019} analyzed \ac{eeg} signals from motor imagery activities without Random Forest, but compared the results with algorithms from Random Forest. In \cite{lazurenkoa2019}, they developed a neural network network strategy to detect EEG signal patterns during motor imagery activities. They compared the results obtained with the Random Forest algorithms, linear discriminant analysis, and linear regression methods considering radial basis functions. The solution presented satisfactory results compared to other algorithms with 82.5$\%$ accuracy in the classification. \cite{meziani2019} used quantile regression and regularization of the L1 norm to estimate the spectrum of the \ac{eeg} signal related to imaging tasks. The study also compared the results with other classification algorithms, including Random Forest. None of the works analyzed in this theoretical framework compared motor imagery \ac{eeg} signals with machine learning. One of the objectives of this paper is to evaluate the performance of a consumer-grade EEG sensor (Mindwave, Neurosky) in an application involving motor imagery.

Consumer-grade and low-cost systems for acquiring \ac{eeg} signals are in BCI research applications \cite{rieiro2019}. The authors of \cite{banu} used the device Mindwave in two applications: controlling the movements of a robot back and forth and controlling electronic devices. In \cite{girase}, the target device was Neurosky, with an accuracy of around $95\%$ in controlling a wheelchair using measures of attention, meditation level, and blinking identification. Still, in the context of wheelchair control using Mindwave, \cite{siswoyo} researchers used neural networks, especially the Backpropagation algorithm, to predict the direction of an electric wheelchair using the \ac{eeg} signal of a person with motor limitations. The results presented indicate a correlation coefficient of 0.92804, showing that the accuracy achieved was satisfactory.


Searching for articles that compare consumer-grade \ac{eeg} sensors with motor imagery-based machine learning algorithms has yet to find published papers. Additionally, Web of Science, Scopus, and PubMed databases did not present research using Random Forest as the primary classification algorithm. Also, most papers used \ac{eeg} data from the \ac{bci} Competition~\cite{BBCI}. This is a database by the Berlin Brain-Computer Interface project, a partnership between the USA and Europe to provide systems that allow a direct dialogue between men and machines. Therefore, instead of recording \ac{eeg} data, most papers use databases. These considerations elucidate the contributions of this paper.

Table \ref{tab:key_papers} summarizes all the research above.

\begin{table}[H]
\caption{Summary of key research papers on EEG signal analysis.\label{tab:key_papers}}
\begin{adjustwidth}{-\extralength}{0cm}
		\newcolumntype{C}{>{\centering\arraybackslash}X}
		\begin{tabularx}{\fulllength}{CCCC}
\toprule
\textbf{Reference}              & \textbf{Motor Imagery} & \textbf{Data Acquisition}                    & \textbf{ML Technique}                                \\ \midrule
Amin et al.~\cite{amin2019} 
      & Yes & Public dataset                         & Convolutional Neural Networks                                                \\
Banu et al.~\cite{banu} & No  & Mindwave                               & -                                                                            \\
Girase et al. ~\cite{girase}  & Yes & Mindwave                               & -                                                                            \\
Lazurenko et al.~\cite{lazurenkoa2019} & Yes & EEG monopolarly from 13 standard leads & Neural Networks                                                              \\
Liu et al.~\cite{liu2017} & Yes & BioSemi                                & Random Forest                                                                \\
\multirow{2}{*}{Meziani et al.~\cite{meziani2019}} & \multirow{2}{*}{Yes}   & \multirow{2}{*}{32 active dry electrode kit} & Random Forest, K-Nearest Neighbors, Neural Networks, \\
                 &     &                                        & Support Vector Machines and Linear Discriminant Analysis \\
Rieiro et al. ~\cite{rieiro2019}  & No  & Mindwave and SOMNOwatch                & -                                                                            \\
Sadiq et al.~\cite{Sadiq2019}  & Yes & Public dataset                         & Support Vector Machines                                                      \\
Siswoyo et al.~\cite{siswoyo} & No  & Neurosky mindset                       & Neural Networks                                                              \\ \bottomrule
\end{tabularx}
\end{adjustwidth}
\end{table}

Therefore, to guide the investigation, we propose to discuss the following questions:

\begin{itemize} 
\item Does \ac{ml} based on random forest algorithm classify different motor tasks (real or imagery)?
\item Does \ac{ml} based on random forest differentiate real and motor imagery in FP1 electrode? 
\item What is the classification performance comparing consumer-graded and research EEG devices?
\item What is the variability between \ac{eeg}'s spectro-temporal and spatial distribution characteristics among subjects?
\end{itemize}

\section{Proposed Solution}
\label{sec_sol}
\begin{figure}[H]
    \includegraphics[width=0.658\textwidth]{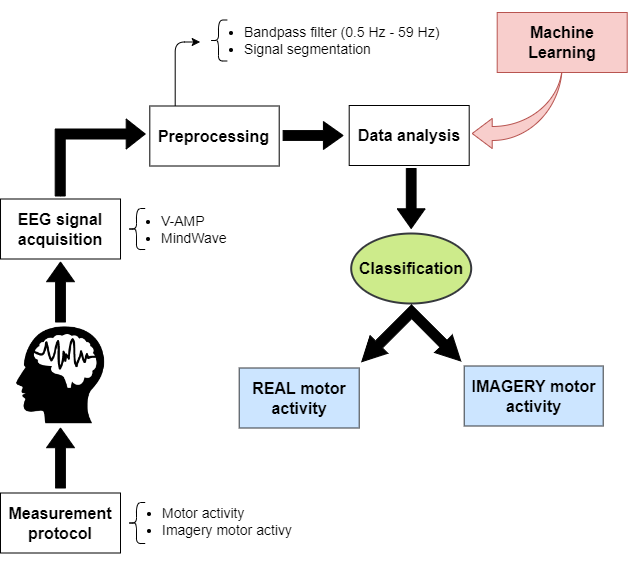}
    \caption{The proposed classification system starts with motor real or imagery tasks. A consumer-graded and a research \ac{eeg} device simultaneously acquire brain activity. The system preprocesses \ac{eeg} signals to enter the machine learning algorithms. The expected result is compared to the trial performed by the subject, and the classification identifies the task, motor real or imagery.}
    \label{fig:diagram_solution}
\end{figure}

In order to be able to distinguish movements based on \ac{eeg} signs, we propose the classification system presented in Figure~\ref{fig:diagram_solution}. 

\subsection{Signal Preprocessing}
\label{subsec_signal_preprocessing}
\ac{eeg} signals are intrinsically noisy and sensitive to other artifacts, like those originating from blinking and muscular activities \cite{teplan2002}. Furthermore, \ac{eeg} has a high temporal resolution (millisecond scale), but a low spatial resolution (centimeter scale), and their placement also influences the quality of the signal. A diffuse signal's spatial resolution introduces an additional challenge to identifying and isolating the \ac{eeg} data task related \cite{craik2019}.

Digital filters attenuate noises efficiently \cite{tese_amilcar}. In this proposed classification system we chose a Butterworth band-pass, 3rd order, 0.5 Hz to 59 Hz, filters powerline hum and keeps brain rhythms (delta to low gamma).

\begin{figure}[H]
\begin{tabular}{cc}
\includegraphics[width=0.48\linewidth]{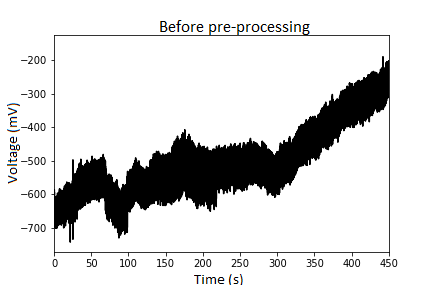}
&\includegraphics[width=0.48\linewidth]{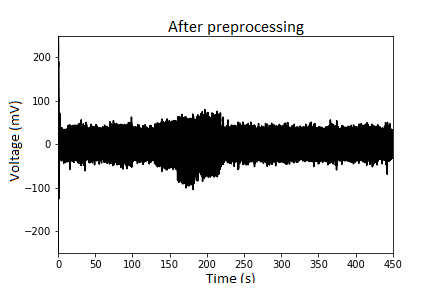}\\
({\bf a})  \ac{eeg} signal before applying the filter.&({\bf b}) \ac{eeg} signal after applying the filter.\\
\end{tabular}
\caption{Illustrative \ac{eeg} signals before and after applying the Butterworth band pass filter. The baseline drift of the signal before preprocessing is evident (\textbf{a}) and after removing the drift (\textbf{b}).}
\label{fig:eeg_filtro}
\end{figure}

Another characteristic of the \ac{eeg} signal is the baseline drift, usually caused by eye movements, deep breaths, and facial muscle movements \cite{Lo2001}. Figure \ref{fig:eeg_filtro} shows the signal before the preprocessing with a noticeable drift (Figure \ref{fig:eeg_filtro}a), and drift filtered out (Figure \ref{fig:eeg_filtro}b).

Participants perform each task for 30 s (motor real or imagery), and the system discards the first 5 s and last 5 s, using 20 s in between, as further explained in Section \ref{sec_setup}.

\subsection{Machine Learning}
\ac{ml} is a set of data analysis method that are able to fit very complex models to a dataset using algorithms some time classified as artificial intelligence. This artificial intelligence algorithms acquires knowledge by extracting patterns from raw data, enabling it to execute complex tasks with minimal human intervention~\cite{goodfellow2016}. 

\ac{ml} algorithms can solve many types of tasks, including classification and regression. In the first one, the algorithm has to specify to which category some input belongs. In the last one, it has to predict a numerical value given some input parameters~\cite{goodfellow2016}. Object recognition and securities' price predictions are examples of, respectively, classification and regression tasks.

Moreover, those algorithms can use three learning processes: supervised, reinforcement, and unsupervised learning. Supervised learning consists of learning from a training set of labeled examples provided by a "supervisor" during a training phase. Each example has a set of inputs and the correct output (or "label"), indicating how the system should behave in that specific case. The goal is for the algorithm to generalize its responses to produce the correct output for sets of inputs not present in the training examples~\cite{sutton2018reinforcement}. 

Reinforcement learning also aims to find the best output to a set of inputs. However, instead of a label, the system provides the algorithm with feedback that shows how good the choice was, but not if it is the best. On the other hand, unsupervised learning comprises finding structures hidden in unlabeled data~\cite{sutton2018reinforcement}.  

This work uses a supervised classification \ac{ml} algorithm called Random Forest. A Random Forest consists of an ensemble of Decision Trees, a non-parametric \ac{ml} algorithm, typically grown using the Classification and Regression Tree method~\cite{breiman2017classification}. For each set of inputs, each Decision Tree generates an output. Then, in the classification problems, the Random Forest chooses the most popular output as its result~\cite{tese_tarciana}.

The fundamental concept of Random Forests is simple but efficient. This algorithm is based on group knowledge. Many relatively uncorrelated models (trees) working together will overcome any individual models. While a tree might converge to the wrong result, most others will likely choose the correct answer. So the group will classify correctly~\cite{russell2009}.

\subsection{Implementation of the Proposed Solution}
This research aims to differentiate a set of motor (real and imagery) activities performed by individuals following an established protocol. This protocol, detailed in~Section \ref{sec_setup}, comprises right and left-hands and right and left ankles movements. The proposed solution must be able to differentiate if the individual imagined or moved a limb and which limb it was.

Before applying the machine learning algorithms, we preprocessed the data, as explained in~Section \ref{subsec_signal_preprocessing}. After, we calculated the statistical moments from order 1 to 10 of the resulting data; those moments serve as features for our \ac{ml} algorithms. Then, we split the data between train, validation, and test sets. We use the first two sets to make a feature selection and a grid search process, defining inputs and parameters of the Random Forests. Both sets division, feature selection and grid search used the \textit{Scikit-Learn} library of Python 3.6.3. 

\begin{figure}[H]
    \includegraphics[width=\textwidth]{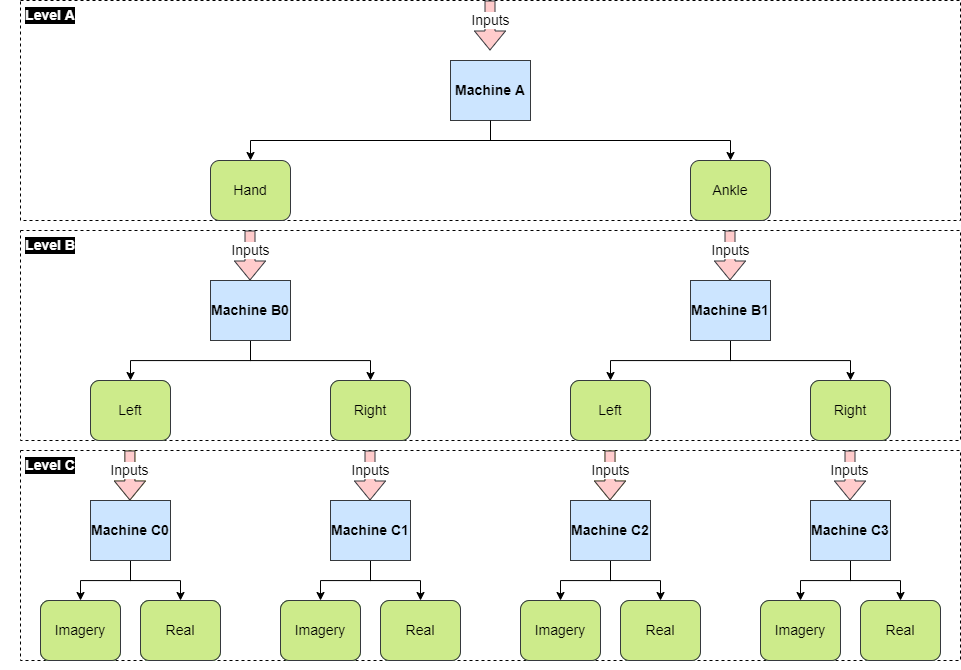}
    \caption{Block diagram of the proposed hierarchical classification architecture}
    \label{fig:algoritmo_maquinas}
\end{figure}

We divide our classification model into three levels: Level A, Level B, and Level C. Each level is responsible for distinguishing a characteristic of the movement, as the literature suggests that the more specialized the machine, the better its results~\cite{russell2009,richards2018}. Each level has the following functionality, as shown in~Figure \ref{fig:algoritmo_maquinas}: 
\begin{itemize}
    \item \textbf{Level A} 
 comprises a single machine responsible for classifying the data according to the region of the movement, i.e., hands or ankles;
    \item \textbf{Level B} contains two machines, one for each region of the movement. The machines at Level B classify the movements between right and left. Considering that, at this point, we have already classified the data by its region, the possible outputs are left-hand, right-hand, left-ankle, and right-ankle movements;
    \item \textbf{Level C} determines whether the subject performed the movement or imagined it. It comprises four machines, each specialized in one possible output of Level B. Therefore, there are eight outputs for Level C: movement of the left-hand, imagery movement of the left-hand, movement of the right-hand, imagery movement of the right-hand, movement of the left ankle, imagery movement of the left ankle, movement of the right ankle, imagery movement of the right ankle.
\end{itemize}

Hence, after Level C, we have the category split intended.

\subsubsection*{Frameworks' Definition}
\begin{figure}[H]
    \includegraphics[width=\textwidth]{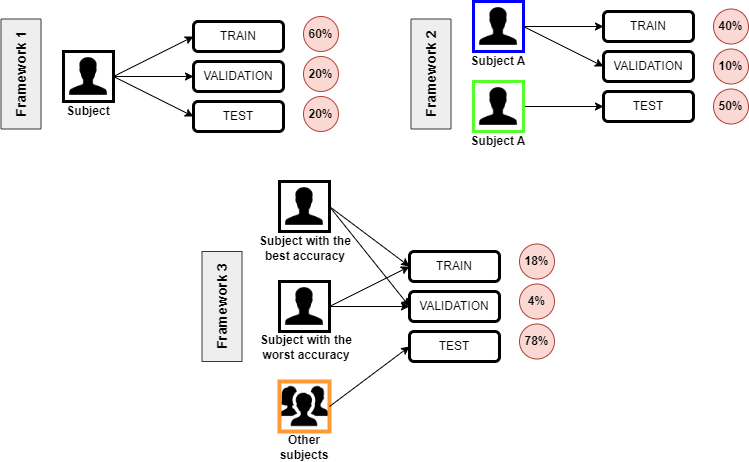}
    \caption{Frameworks used in the research.}
    \label{fig:frameworks}
\end{figure}

Interpersonal variability is a challenge to \ac{eeg} signals classification, especially when motor imagery is involved, as people tend to imagine or do things in their particular way. Considering that, we defined three frameworks to verify the accuracy of our proposed solution. Each framework splits the data between the train, test, and validation subsets in a particular way, allowing us to evaluate the correlation between \ac{eeg} signals from different subjects, as shown in~Figure \ref{fig:frameworks}. Therefore, the proposed solution also aims to identify the difficulty level when we consider different data from different subjects.

The Frameworks differ from one another regarding their way of defining the data from the train, validation, and test subsets, as further explained:
\begin{itemize}
    \item \textbf{Framework 1}: used the same data collection from the same subject to train, validate, and test;
    \item \textbf{Framework 2}: considered two data collections from the same subject and used one of them to train and validate and the other for testing;
    \item \textbf{Framework 3}: to test and validate, we used data collections from the two subjects with the best and the worst accuracy in Framework 1's classification. Then, we test it with data collections of the remaining seven subjects.
\end{itemize}

Moreover, concerning the size of the train, test, and validation subsets, the frameworks do as follows:
\begin{itemize}
    \item \textbf{Framework 1}: 60\% of data for train, 20\% for validation, and 20\% for test subsets;
    \item \textbf{Framework 2}: 40\% of data for train, 10\% for validation, and 50\% for test subsets;
    \item \textbf{Framework 3}: 18\% of data for train, 4\% for validation, and 78\% for test subsets.
\end{itemize}

As the increase in the data variability tends to cause a decrease in accuracy, we expect Framework 1 to have the best results, followed by Framework 2 and then Framework 3. These results would show the uniqueness of each person's \ac{eeg} signals, even when performing the same research protocol as others.

Furthermore,  it is essential to notice that the choices made for each framework also have implications for their applicability. In Framework 1, the low variability implies the need to train its machines for each use of the subject. Framework 2, however, only needs to train its machines once for each subject. At least, Framework 3, due to its high variability, does not need a training process for each subject or each use. Its machines only need to be trained once to be appropriate for the use of anyone at any time. Thus, there is a clear trade-off between an easier classification and a more straightforward application.

\section{Experimental Setup and Measurement Protocol}
\label{sec_setup}

\subsection {Experimental Setup}

\begin{figure}[H]
		\includegraphics[height=0.4\textheight]{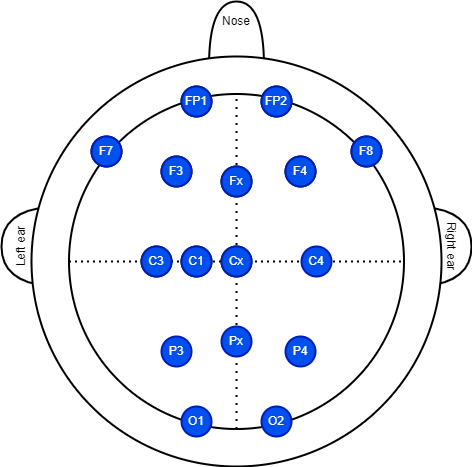}
		\caption{Position of the electrodes we used in the \ac{eeg} signal aquisition with V-AMP.}
		\label{fig:vamp_eletrodos}
\end{figure}

The proposal is to analyze EEG signals of individuals submitted to motor imagery tasks. The protocol used two devices: a 16-channel \ac{eeg} system (V-AMP amplifier) and a single-channel consumer-grade EEG (Mindwave). The V-AMP amplifier (Brain Products GmbH)~\cite{Brain_Products} uses an easycap for positioning 16 active EEG electrodes, using conductive gel to decrease impedance. Figure \ref{fig:vamp_eletrodos} shows the positioning of the 16 electrodes based on 10--20 electrode system.

\begin{figure}[H]
		\includegraphics[height=0.3\textheight]{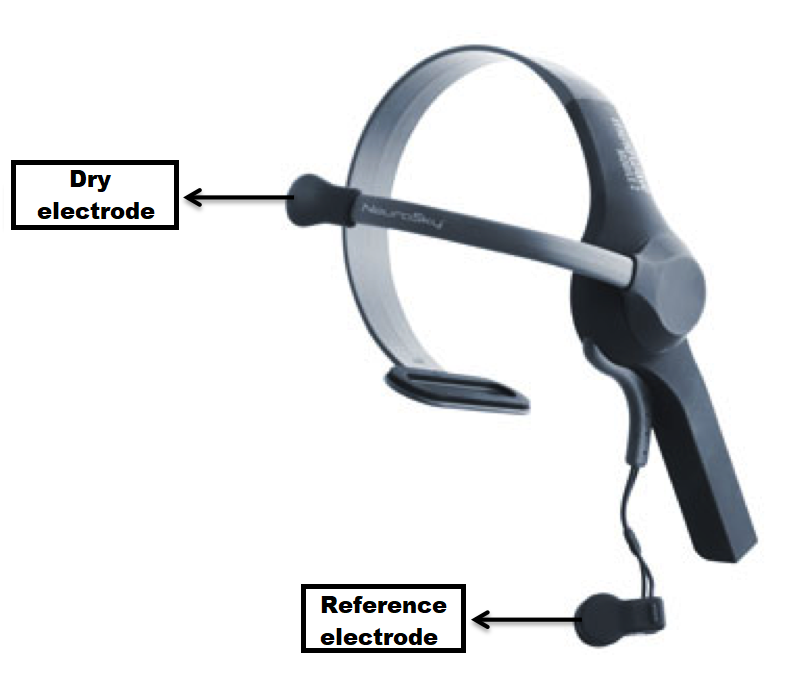}
		\caption{A Mindwave device from the company \textit{Neurosky}. 
 Adapted from~\cite{mindwave}.}	
		\label{fig:mindwave}
\end{figure}

The Mindwave \cite{mindwave}, depicted in Figure \ref{fig:mindwave}, is a device developed by Neurosky. It consists of a single dry \ac{eeg} electrode placed in the forehead, above the eyebrow, and it features an ear lobe. It uses Bluetooth to to stream the data.

The software used to acquire the signals from both sensors was OpenVIBE \cite{OpenVibe}. The data were collected simultaneously. Therefore, it was possible to analyze the same dataset from the two sensors at the same time. The sampling frequency for both is 512~Hz, almost nine times higher than the highest desired signal frequency (59~Hz, see Section \ref{sec_sol}), respecting Nyquist criterion.

\begin{figure}[H]
		\includegraphics[height=0.25\textheight]{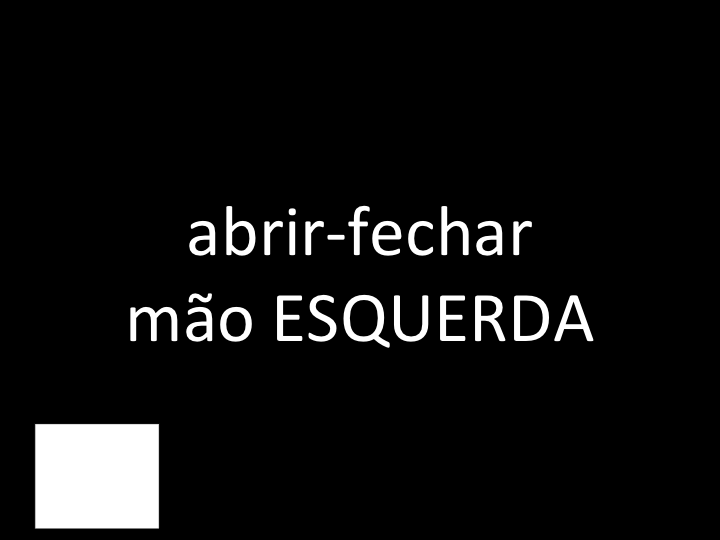}
		\caption{Image presented to the subject with the instruction “open and close the left-hand” in Portuguese. We use the white square in the bottom left to activate the light sensor and indicate the beginning and end of each activity.}
		\label{fig:imagem_tarefa}
\end{figure}

\begin{figure}[H]
		\includegraphics[height=0.30\textheight]{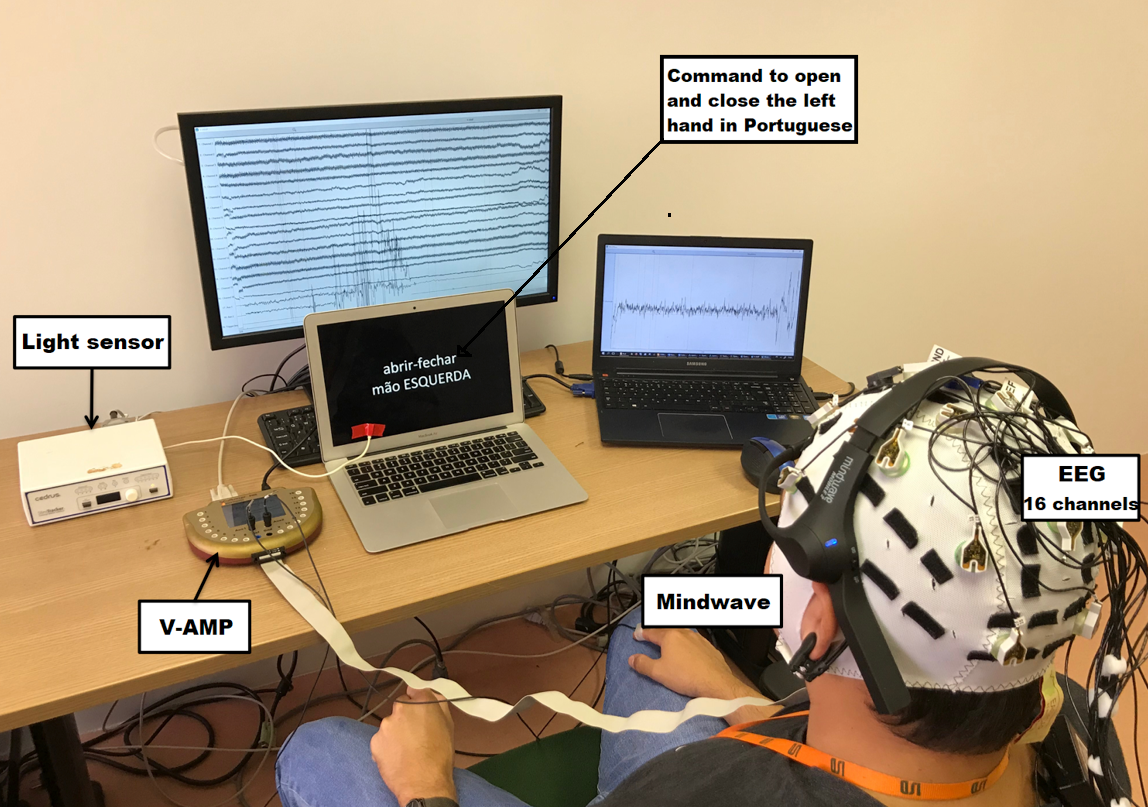}
		\caption{Experimental setup for \ac{eeg} recording using V-amp and Mindwave, light sensor to synchronize instruction presentation with data recorded.}
		\label{fig:setup}
\end{figure}

Another hardware we added to the system was the StimTracker~\cite{Cedrus} manufactured by Cedrus. This equipment worked as a light sensor. The image presented to the candidate (shown in Figure \ref{fig:imagem_tarefa}) contains  a white square that appears for 0.5~seconds at the beginning and end of each activity. The light sensor is connected to the screen and identifies when the white square appears. The light sensor signal works in parallel with the \ac{eeg} sensor signals in the OpenVIBE software. Therefore, it is possible to synchronize the \ac{eeg} signal when the task starts and ends, eliminating the dependence on manual time stamping. Figure \ref{fig:setup} shows the equipment layout and collection scenario.

\subsection{Measurement Protocol}

This project was approved by the Local Ethical Committee (CAAE: 79649717.0.0000.5292). The measurement protocol considered 14 healthy participants (10 male and 04 female, aged between 20 and 30 years). The subjects were seated comfortably in a chair and placed their hands on their legs, as illustrated in Figure \ref{fig:setup}. They executed the sequence of activities presented through a computer screen positioned in front of them, as shown in Figure \ref{fig:imagem_tarefa}. The set of activities comprised motor and imaginary tasks of the hands and ankles, according to the following task orientation:

\begin{itemize}  
	\item Imagine opening and closing the right-hand;  
	\item Open and close the right-hand;  
	\item Imagine opening and closing the left-hand;  
	\item Open and close left-hand;  
	\item Imagine flexing the right ankle;  
	\item Flex the right ankle;  
	\item Imagine flexing the left ankle ; 
	 \item Flex the left ankle.  
\end{itemize}

Each task lasted 30 seconds without interval. Each subject performed the protocol twice, resting a few minutes between the recordings. OpenVIBE recorded \ac{eeg} data from V-amp, Mindwave, and light sensor. \ac{eeg} signals from V-amp and Mindwave are displayed in real-time on two monitors.

\subsubsection*{Data Selection}

Data exclusion attended the following criteria:

\begin{itemize}  
	\item Participants who did not follow the task instruction;
	\item Data with missing synchronization trigger;
	\item Low battery of the Mindwave, which impaired data streaming. 
\end{itemize}
	
Data from 14 participants were recorded, 5 were excluded, and 9 satisfied the system analysis conditions to guarantee reliable results.

\section{Results and Discussions}
\label{sec_results}

\subsection{Preliminary Analysis}

\begin{figure}[H]
		\includegraphics[height=0.4\textheight]{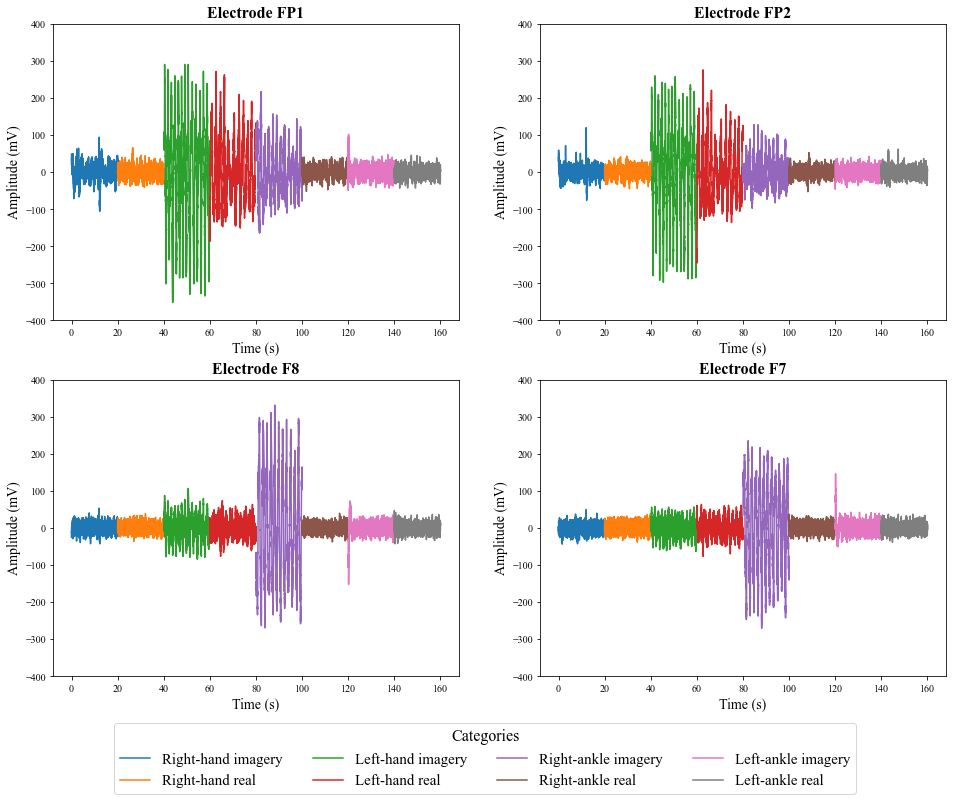}
		\caption{\ac{eeg} signals captured by the first 4 channels of the sensor V-AMP for \textbf{Subject 01}.}
		\label{fig:vamp1_power}
\end{figure}

\begin{figure}[H]
		\includegraphics[height=0.4\textheight]{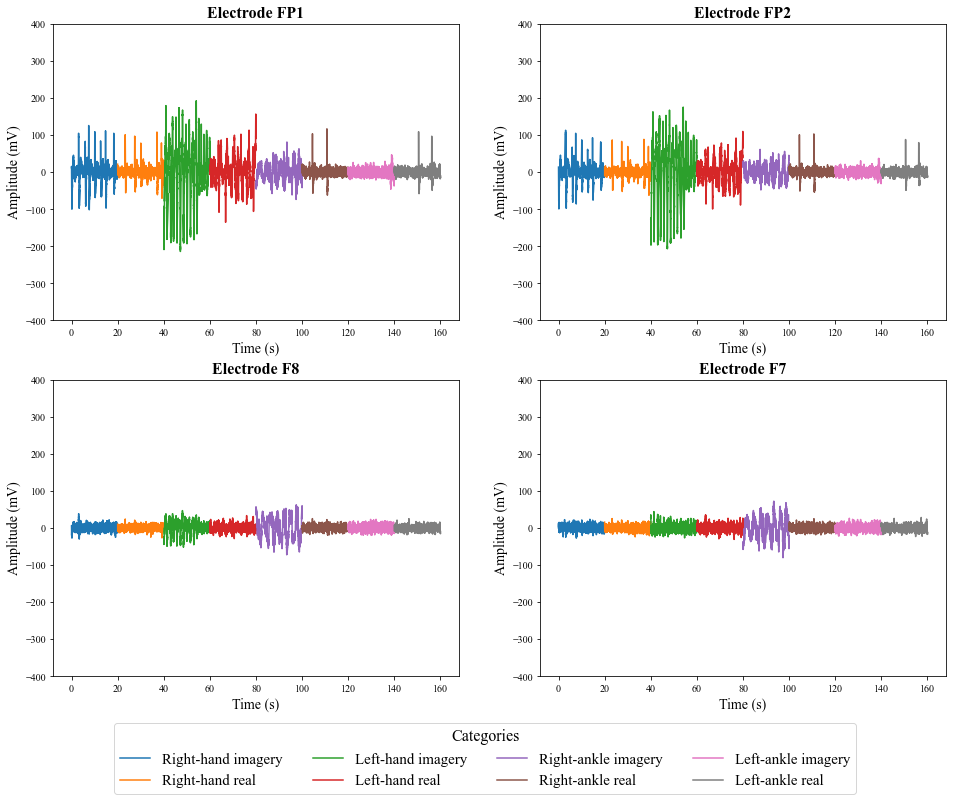}
		\caption{\ac{eeg} signals captured by the first 4 channels of the sensor V-AMP for \textbf{Subject 02}.}
		\label{fig:vamp2_power}
\end{figure}

\begin{figure}[H]
		\includegraphics[height=0.4\textheight]{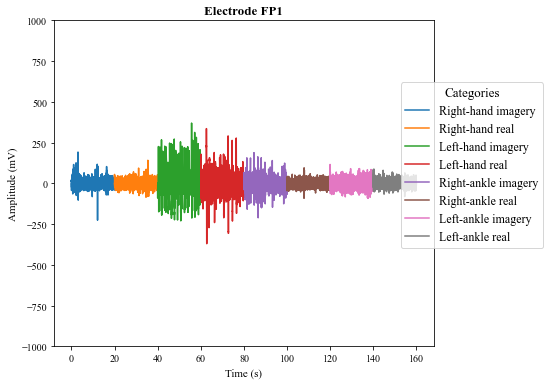}
		\caption{\ac{eeg} signals captured by the sensor Mindwave for \textbf{Subject 01}.}
		\label{fig:mw1_power}
\end{figure}

\begin{figure}[H]
		\includegraphics[height=0.4\textheight]{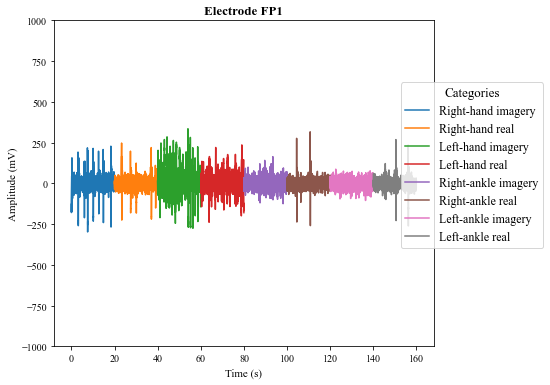}
		\caption{\ac{eeg} signals captured by the sensor Mindwave for \textbf{Subject 02}.}
		\label{fig:mw2_power}
\end{figure}

Figure \ref{fig:vamp1_power,fig:vamp2_power} illustrate the signals captured by the first four channels of the V-AMP sensor for two different subjects. Similarly, Figure \ref{fig:mw1_power,fig:mw2_power} show the signals of the same individuals, in the same collection, by the Mindwave sensor. The graphic presentation of these signals in the referred figures considers the signal strength in time, and a specific color identifies each task. Therefore, it is possible to analyze in time how \ac{eeg} signal behaves for each person during each activity. The first analysis is that even performing identical tasks, following the same protocol, each person presents unique characteristics in the signal data set.

\begin{figure}[h]
		\includegraphics[height=0.4\textheight]{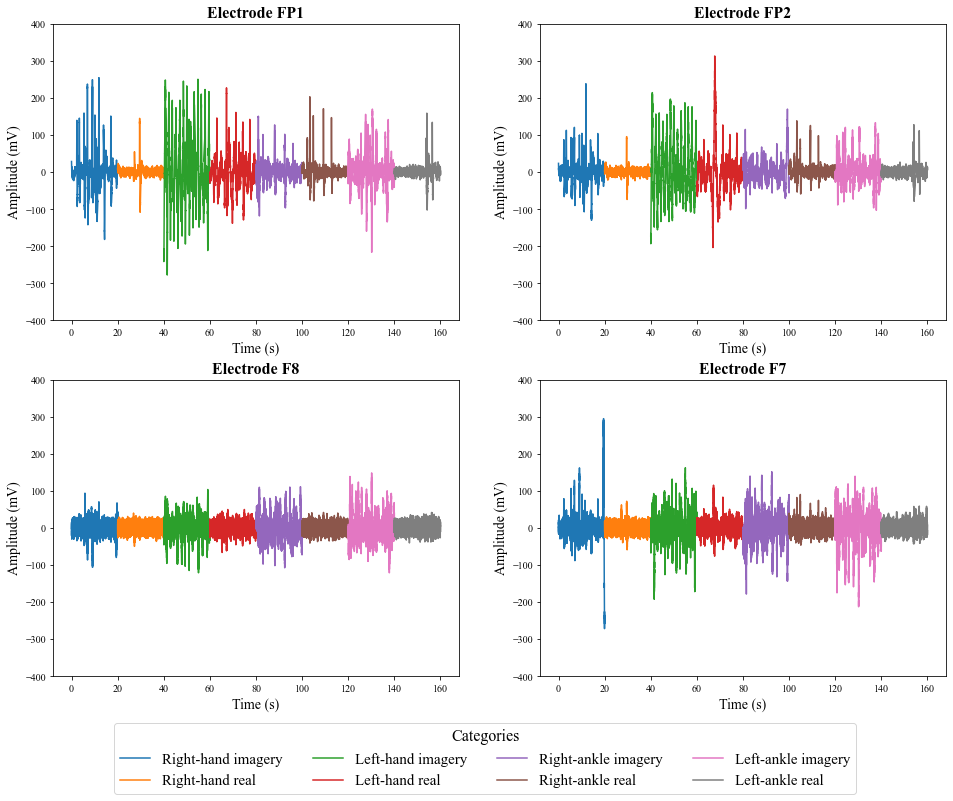}
		\caption{\ac{eeg} signals captured by the first 4 channels of the sensor V-AMP for the first collect of a subject.}
		\label{fig:mesmo_individuo1}
\end{figure}

\begin{figure}[h]
	   	\includegraphics[height=0.4\textheight]{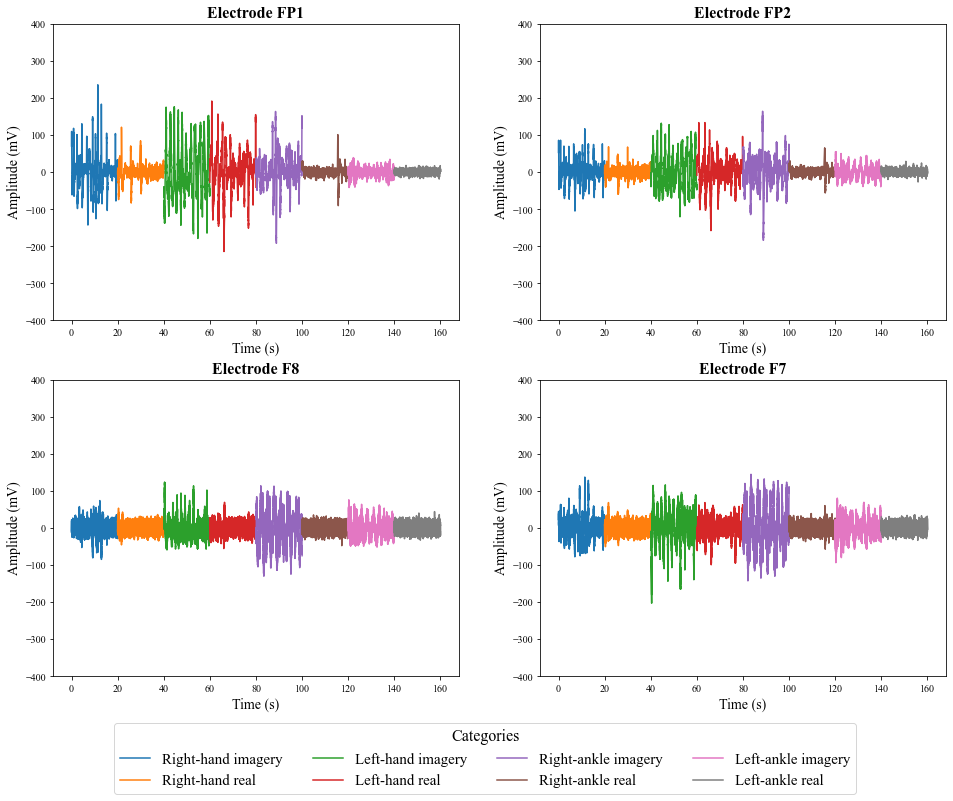}
	   	\caption{\ac{eeg} signals captured by the first 4 channels of the sensor V-AMP for the second collect of a subject.}
		\label{fig:mesmo_individuo2}
\end{figure}

The results also show that in addition to the interpersonal variation, the \ac{eeg} signals present variations for the same person, even when that person executes the same protocol. The electroencephalographic signal does not have a defined visual pattern for a given action. If the same person performs an activity at a specific time and then repeats the activity at another time, the signal will probably present another visual pattern. \mbox{Figures \ref{fig:mesmo_individuo1} and \ref{fig:mesmo_individuo2}} show the signs of the same individual in two different collections with the first four channels of V-AMP. In the images, regardless of which sensor is responsible for the acquisition, the signals are visually different for each recording, and highlight the need of a \ac{ml} classifier.

\subsection{Classifier Results Analysis}

\begin{figure}[h]
\begin{tabular}{cc}
\includegraphics[width=0.4\linewidth]{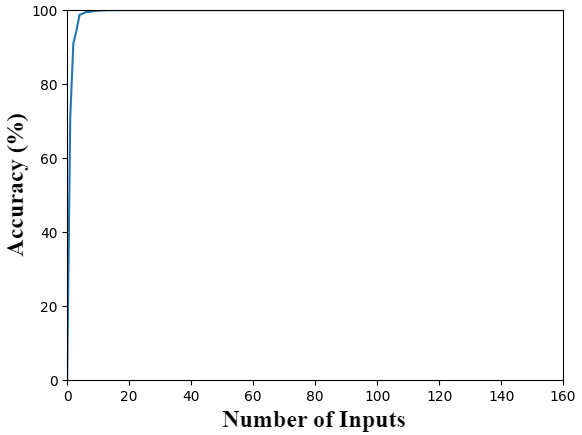}  
&\includegraphics[width=0.4\linewidth]{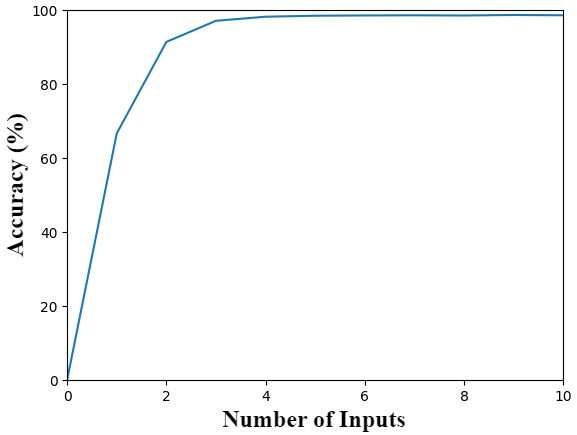}  \\
({\bf a})  V-AMP.&({\bf b}) Mindwave.\\
\end{tabular}
\caption{Accuracy of a feature selection for a machine in Level A.}
\label{fig:featureselection}
\end{figure}

Data processing used machine learning algorithms and Random Forest as a classification tool. One of the steps developed was Feature Selection, which generates better output in the \ac{ml} algorithm. Figure \ref{fig:featureselection} contemplates the result of the Feature Selection for a given subject, considering the machine learning process at Level A, which classifies into hands or ankles. A controlled amount of inputs in the classification process optimizes the algorithm's processing, which requires a smaller set of mathematical operations. The results show that the machine needs to use around $5$ inputs to reach a satisfactory accuracy, around $100\%$. Both devices, V-AMP and Mindwave, presented this pattern.

As shown in the diagram in Figure \ref{fig:algoritmo_maquinas}, the project comprises the implementation of seven machines learning for the analysis and classification of \ac{eeg} signals. The machines are divided into levels and specialize in certain classifications. Each machine, at a specific level, implements Feature Selection; this implies that some channels and statistics are more relevant to the classification process for each machine. Table~\ref{tab:feature_selection_vamp} presents the result of the Feature Selection of one of the research participants for V-AMP data.

\begin{table}[H]
\caption{\label{tab:feature_selection_vamp} Feature Selection Results for V-AMP data.}
\newcolumntype{C}{>{\centering\arraybackslash}X}
		\begin{tabularx}{\textwidth}{CCC}
\toprule
\textbf{Machine} & \textbf{Moments} & \textbf{Channels} \\ 
\midrule
A & 2 and 4 & Fp2 and F4 \\  
B0 & 6 & C3 and O2 \\ 
B1 & 2, 4 and 6 & F3 and P3 \\
C0 & 3, 5, 7, 8, 9 and 10 & Cz, C4 and Pz \\
C1 & 4, 6, 8 and 10 & F3 and F4 \\ 
C2 & 8 & O2 \\ 
C3 & 2, 4, 6, 8 and 10 & P3, Pz, P4, O1 and O2 \\ 
\bottomrule
\end{tabularx}
\end{table}

Mindwave is a single-channel sensor, so the Feature Selection process only considers the moments calculated. Table~\ref{tab:feature_selection_mw} shows the results of the same individual for the Mindwave data.

\begin{table}[H]
\caption{Feature Selection Results for Mindwave data.\label{tab:feature_selection_mw}}
\newcolumntype{C}{>{\centering\arraybackslash}X}
		\begin{tabularx}{\textwidth}{CCC}
\toprule
\textbf{Machine} & \textbf{Moments} \\ 
\midrule
A & 2, 3, 4, 5, 6 and 7 \\ 
B0 & 2 e 4 \\ 
B1 & 2, 3, 4, and 6 \\ 
C0 & 2, 3, 4, 5, 6, 7 and 8 \\ 
C1 & 2, 4, 5, 7, and 9 \\ 
C2 & 6 and 8 \\ 
C3 & 2, 4, 6, 7 and 8 \\ 
\bottomrule
\end{tabularx}
\end{table}

We decided to consider the best and worst results to generalize the data exposure of the nine subjects. The metric used was the average of the accuracies of each activity, resulting from the confusion matrix. In other words, the results presented are for the individuals who obtained the highest and lowest average in the activities classification.

\subsubsection{Framework 1}
\label{subsection:frame1}

\begin{figure}[H]
\begin{adjustwidth}{-\extralength}{0cm}
\centering
		\includegraphics[height=0.25\textheight]{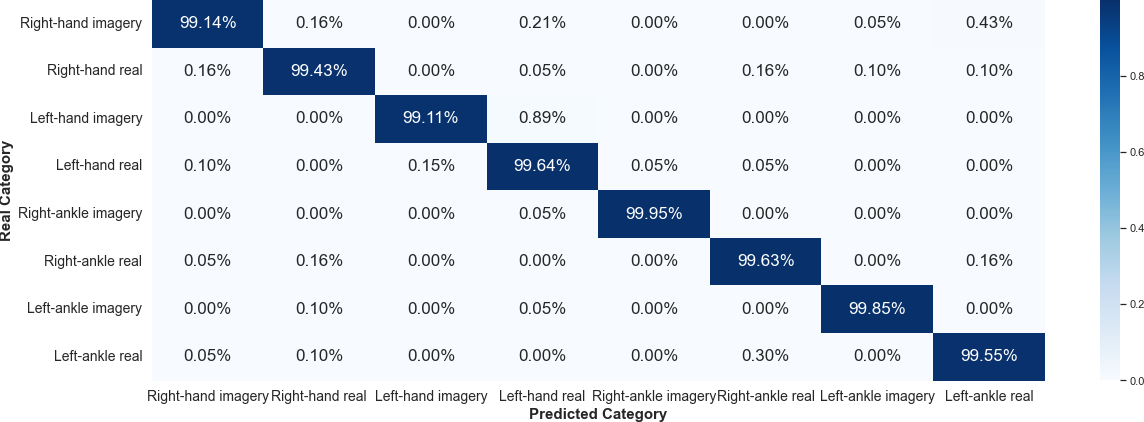}
\end{adjustwidth}
		\caption{Confusion matrix of Framework 1 using the sensor V-AMP. Subject with the best score.}
		\label{fig:fw1_melhor_vamp}
\end{figure}

\begin{figure}[H]
\begin{adjustwidth}{-\extralength}{0cm}
\centering
		\includegraphics[height=0.25\textheight]{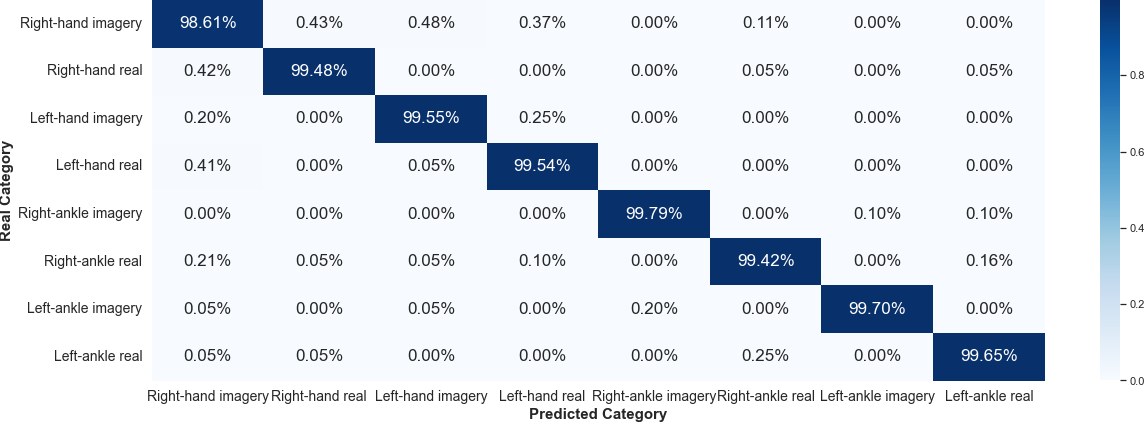}
\end{adjustwidth}
		\caption{Confusion matrix of Framework 1 using the sensor V-AMP. Subject with the worst score.}
		\label{fig:fw1_pior_vamp}
\end{figure}

\begin{figure}[H]
\begin{adjustwidth}{-\extralength}{0cm}
\centering
		\includegraphics[height=0.25\textheight]{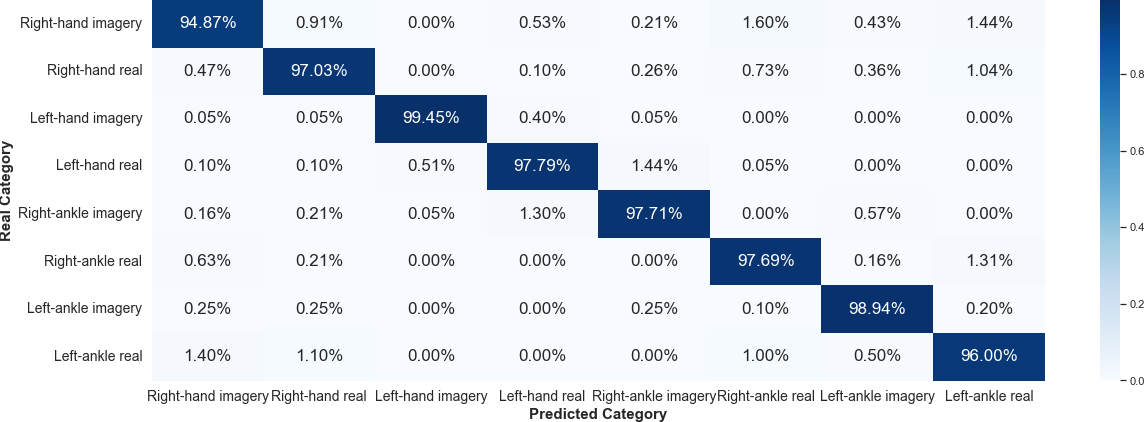}
\end{adjustwidth}
		\caption{Confusion matrix of Framework 1 using the sensor Mindwave. Subject with the best score.}
		\label{fig:fw1_melhor_mw}
\end{figure}

\begin{figure}[H]
\begin{adjustwidth}{-\extralength}{0cm}
\centering
		\includegraphics[height=0.25\textheight]{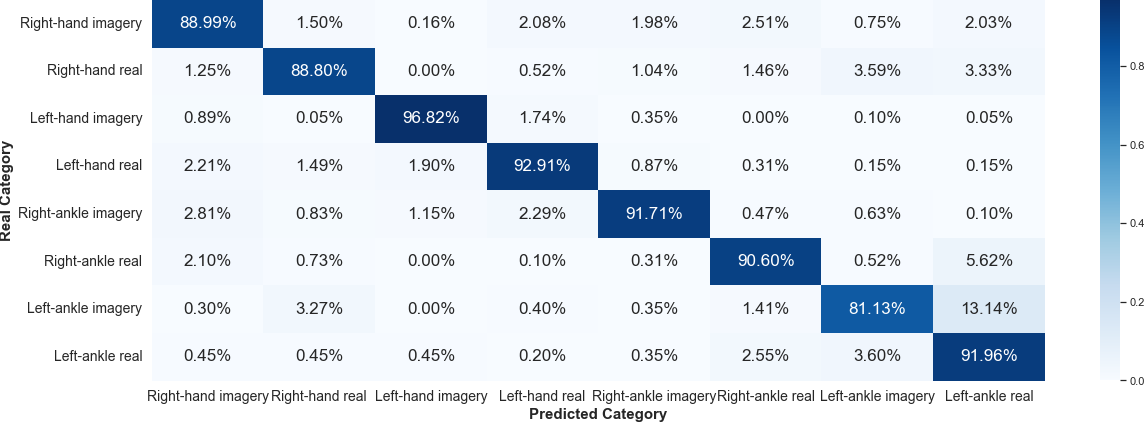}
\end{adjustwidth}
		\caption{Confusion matrix of Framework 1 using the sensor Mindwave. Subject with the worst score.}
		\label{fig:fw1_pior_mw}
\end{figure}

Framework $1$ considered data from the same individual for all stages of the algorithm: training, validation, and testing. Figures \ref{fig:fw1_melhor_vamp} and \ref{fig:fw1_pior_vamp} show the results, via a confusion matrix, for Framework $1$ considering V-AMP data for the individual with best and worst performance, respectively. Figures \ref{fig:fw1_melhor_mw} and \ref{fig:fw1_pior_mw} present the results of the data collected by Mindwave for the same Framework, also considering the best and worst performance.

We expected that this Framework would present a satisfactory result, as this case does not deal with two problems of \ac{eeg} signal analysis: interpersonal variability and collection of the same individual at different times. Comparing the results of V-AMP with Mindwave indicates that even Mindwave is a simple sensor with a dry electrode and low cost, it can provide a high accuracy rate in the classification process.

Another fact exposed by the results is the stability of the accuracy rate between the worst and the best individual, something expected since the algorithm's execution happens independently for each person.

\subsubsection{Framework 2}
\label{subsection:frame2}

\begin{figure}[H]
\begin{adjustwidth}{-\extralength}{0cm}
\centering
		\includegraphics[height=0.25\textheight]{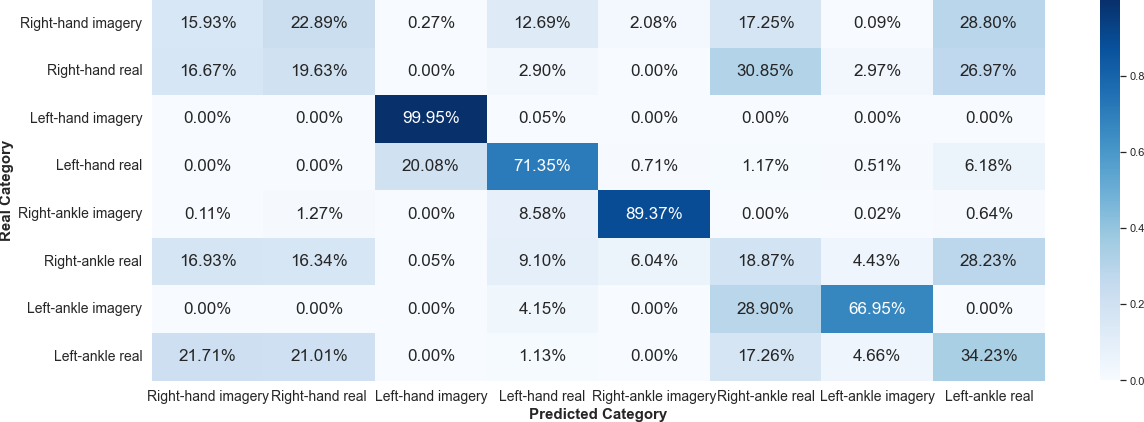}
\end{adjustwidth}
		\caption{Confusion matrix of Framework 2 using the sensor V-AMP. Subject with the best score.}
		\label{fig:fw2_melhor_vamp}
\end{figure}

\begin{figure}[H]
\begin{adjustwidth}{-\extralength}{0cm}
\centering
		\includegraphics[height=0.25\textheight]{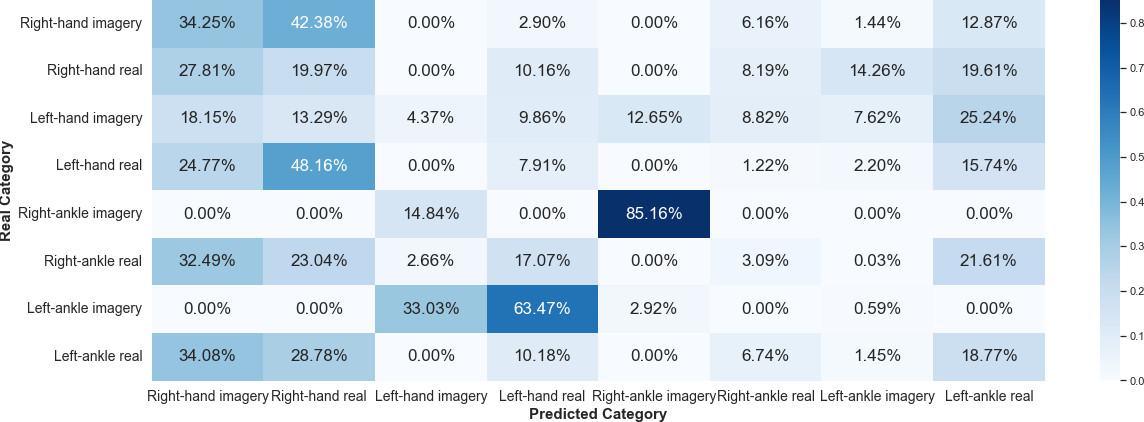}
\end{adjustwidth}
		\caption{Confusion matrix of Framework 2 using the sensor V-AMP. Subject with the worst score.}
		\label{fig:fw2_pior_vamp}
\end{figure}

\begin{figure}[H]
\begin{adjustwidth}{-\extralength}{0cm}
\centering
		\includegraphics[height=0.25\textheight]{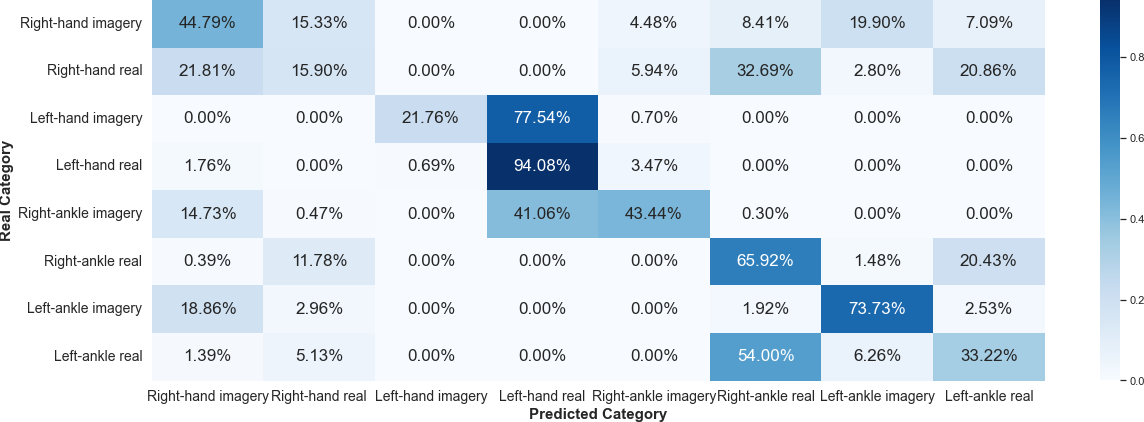}
\end{adjustwidth}
		\caption{Confusion matrix of Framework 2 using the sensor Mindwave. Subject with the best score.}
		\label{fig:fw2_melhor_mw}
\end{figure}

\begin{figure}[H]
\begin{adjustwidth}{-\extralength}{0cm}
\centering
		\includegraphics[height=0.25\textheight]{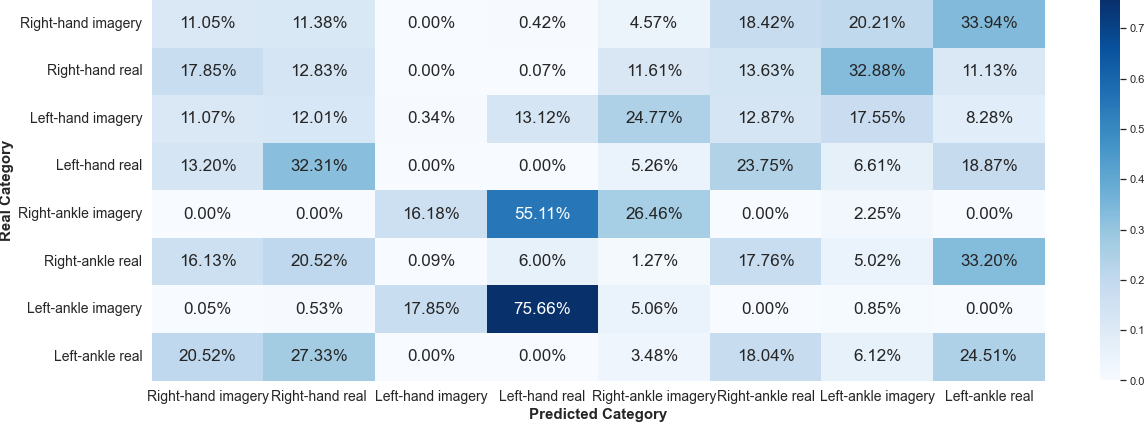}
\end{adjustwidth}
		\caption{Confusion matrix of Framework 2 using the sensor Mindwave. Subject with the worst score.}
		\label{fig:fw2_pior_mw}
\end{figure}

Framework 2 uses data from the same subject considering different collections. The results of V-AMP are shown in Figures \ref{fig:fw2_melhor_vamp} and \ref{fig:fw2_pior_vamp}, and those of Mindwave in \mbox{Figures \ref{fig:fw2_melhor_mw} and \ref {fig:fw2_pior_mw}}.

It is noticeable that compared to Framework 1, the sorting ability of the algorithm for Framework 2 has dropped considerably. This result reaffirms that there is still a significant problem in working with \ac{eeg} data from different temporal experiences, implying that the classification method developed was inefficient in processing \ac{eeg} data collected at different times.

\subsubsection{Framework 3}
\label{subsection:frame3}

\begin{figure}[H]
\begin{adjustwidth}{-\extralength}{0cm}
\centering
		\includegraphics[height=0.25\textheight]{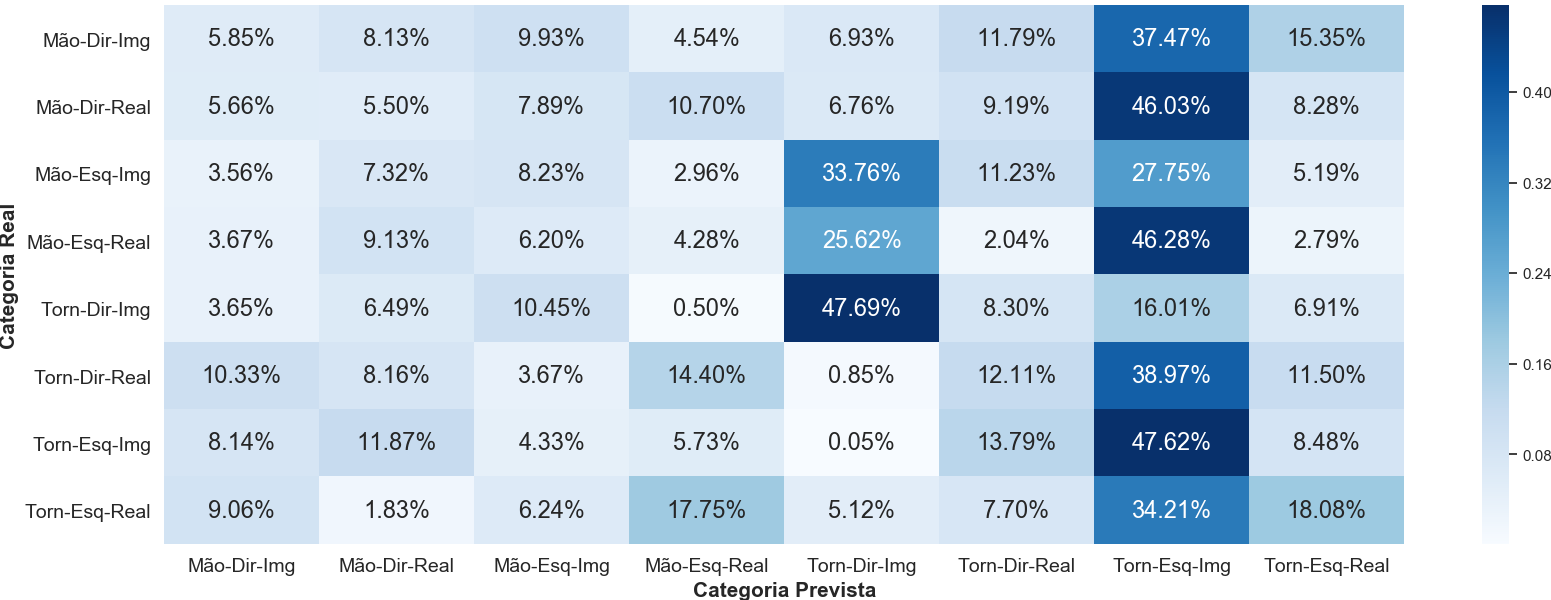}
\end{adjustwidth}
		\caption{Confusion matrix of Framework 3 using the sensor V-AMP.} 
		\label{fig:fw3_vamp}
\end{figure}

\begin{figure}[H]
\begin{adjustwidth}{-\extralength}{0cm}
\centering
		\includegraphics[height=0.25\textheight]{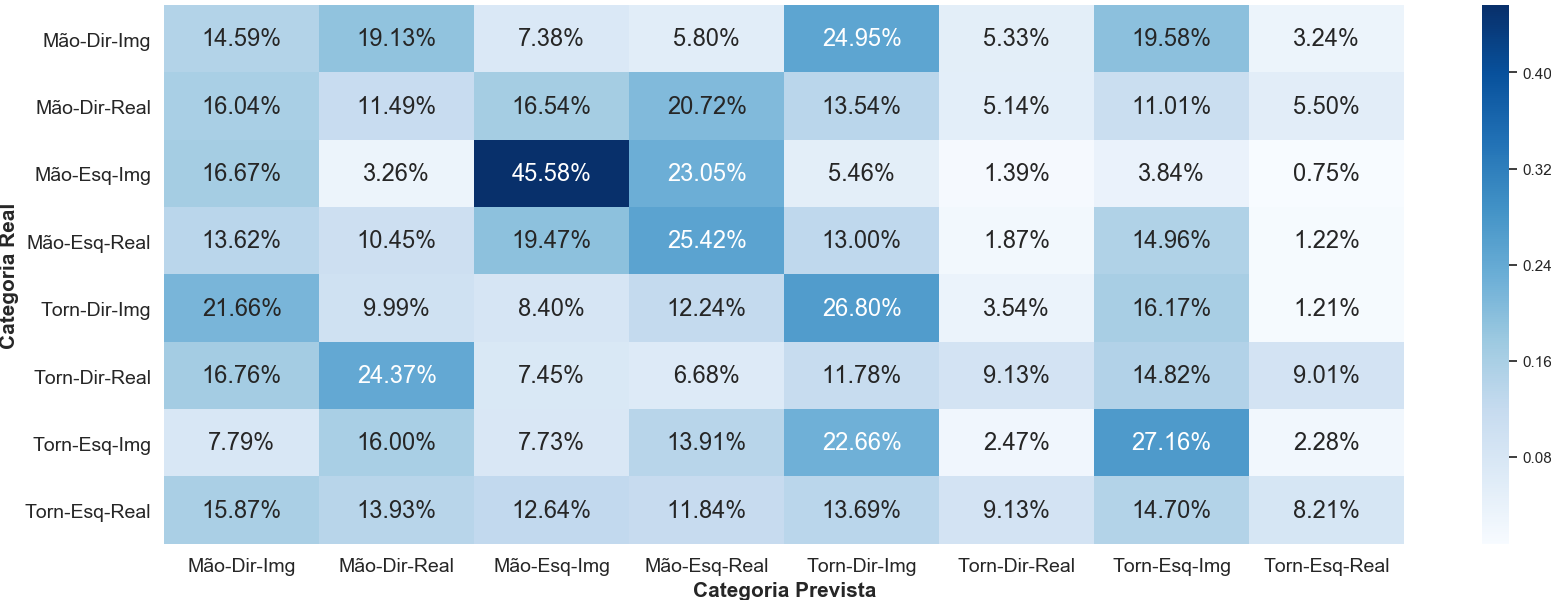}
\end{adjustwidth}
		\caption{Confusion matrix of Framework 3 using the sensor Mindwave.} 
		\label{fig:fw3_mw}
\end{figure}

Framework 3 seeks to increase the level of difficulty of classification. In addition, it aims to attest to the problems already highlighted by other research groups in the area, which point out that the extraction of characteristics from the \ac{eeg} signals is a complex process due to the interpersonal variability of the signals. The result for V-AMP data is shown in Figure \ref{fig:fw3_vamp}, and for Mindwave in Figure \ref{fig:fw3_mw}. For these graphs, we calculated the average confusion matrix among all individuals.

For this Framework, the performance of the sensors V-AMP and Mindwave were very different and did not follow any logical line. The best but unsatisfactory classification with the V-AMP data was for the imagery left ankle movement. With the data from Mindwave, the image movement of the left-hand was the best.

\subsection{Comparison Between Sensors}

The results presented in each framework also indicate that even with different hardware configurations, the devices used in this research could reproduce, in general, consistent results.V-amp as a research device presents superior data quality, and Mindwave even with channel limitations can support specific research applications.

\section{Conclusions and Final Remarks}
\label{sec_conclusions}

The methodology developed in this paper considered implementing a machine learning algorithm with the Random Forest technique to classify real and imagery activities. The set of solutions used converged to satisfactory results. 

Both devices, Mindwave and V-AMP, can be used to acquire electroencephalographic signals. However, Mindwave's data restricts the classification capacity of machine learning for Frameworks 2 and 3. Nevertheless, Mindwave has lower performance than V-AMP since this device was initially conceived for game application and has only one dry electrode positioned in the forehead. Furthermore, more experiments are needed to test the Mindwave applicability in \ac{bci} field. Also, the signal quality of the Mindwave sensor might need to be further improved to reduce artifacts. 

The implemented algorithm is efficient in distinguishing and classifying motor and imagery activities, especially when the data come from the same subject for the same collection, i.e., Framework 1. The results of Frameworks 2 and 3 indicate that dealing with this type of signal is complex when it comes to collections of the same individual at different times and collections of different individuals at different times. Considering this last point, one of the most significant difficulties is the interpersonal variability of \ac{eeg} data. However, the proposed algorithm is helpful in \acp{bci} applications where it is possible to perform training and validation processes in each use.

The answer to the questions in the Section \ref{sec_works} are:

\begin{itemize} 

    \item Does \ac{ml} based on random forest algorithm classify different motor tasks (real or imagery)?
    \begin{quote}
        This paper showed that the \ac{eeg} signals have different characteristics, considering the same subject for a given set of activities. The signal is variable over time, and its characterization process is complex. Nevertheless, \ac{ml} based on random forest, such as the algorithm implemented in this work, is efficient in classifying these signals, as the results presented in Framework 1.
    \end{quote} 
    
    \item Does \ac{ml} based on random forest differentiate real and motor imagery in FP1 electrode? 
    \begin{quote}
         It achieved satisfactory performance for Framework 1, with accuracies above $94\%$. However, for the other frameworks, the classification was not efficient. Therefore, these results indicate that FP1 position can be explored in research that uses the exact data for the entire training and classification process, considering the implementation of machine learning algorithms.
    \end{quote}
    
    \item What is the classification performance comparing consumer-graded and research EEG devices? 
    \begin{quote}
         Mindwave has a lower performance when compared to V-AMP. However, it was expected that Mindwave would present lower classification accuracy. Nevertheless, results for Framework 1 supports that a consumer-graded \ac{eeg} could be used in applications involving \ac{eeg} data classification.
    \end{quote}
    
    \item What is the variability between \ac{eeg}'s spectro-temporal and spatial distribution characteristics among subjects?
    \begin{quote}
        The results showed that \ac{eeg} data presents temporal variability among the subjects. The best result achieved with the proposed methodology and algorithm was for Framework 1, which considered data from the same individual for the same task. However, this fact does not necessarily indicate that this is the only possibility of classification. There are projections of improvements in the algorithm that can provide the classification process with the ability to identify statistics in the signal of tasks performed by different people—suggestions for further research.

    \end{quote}
    \end{itemize}

\vspace{6pt} 



\authorcontributions{Conceptualization, T.C.B.G and T.N.; methodology, T.C.B.G and T.N..; software, T.C.B.G and T.N.; validation, A.M.M., E.M., and V.A.S.J.; formal analysis, T.C.B.G., T.N., A.M.M., E.M., and V.A.S.J.; investigation, T.C.B.G and T.N.; resources, T.C.B.G and T.N.; data curation, T.C.B.G and T.N.; writing—original draft preparation,
T.C.B.G and T.N.; writing—review and editing, T.C.B.G., T.N., A.M.M., E.M., and V.A.S.J.; visualization, T.C.B.G and T.N.; supervision, A.M.M., E.M., and V.A.S.J.; project administration, T.C.B.G and T.N.; funding acquisition, A.M.M., E.M., and V.A.S.J. All authors have read and agreed to the published version of the manuscript.}

\funding{This study was financed in part by the Coordena\c{c}\~{a}o de Aperfei\c{c}oamento de Pessoal de N\'{i}vel Superior - Brasil (CAPES) - Finance Code 001.}

\institutionalreview{The study was conducted in accordance with the Declaration of Helsinki, and approved by the Ethics Committee of Plataforma Brasil (CAAE: 79649717.0.0000.5292), responsible researcher Edgard Morya.}

 \informedconsent{Informed consent was obtained from all subjects involved in the study.}

\dataavailability{The data of this study are available from the corresponding author upon reasonable request.}


\conflictsofinterest{The authors declare no conflicts of interest.} 

\begin{adjustwidth}{-\extralength}{0cm}

\reftitle{References}




\end{adjustwidth}
\end{document}